\documentclass[12pt,preprint]{aastex}
\slugcomment{}

\shorttitle{Treta-Atomic Carbon in the Space $?$}
\shortauthors{Cernicharo et al.}

\begin{document}

\title{A new infrared band in Interstellar and Circumstellar Clouds:
C$_4$ or C$_4$H$?$ \\}

\author{Jos\'e Cernicharo\altaffilmark{1} and Javier R. Goicoechea}
\affil{Instituto de Estructura de la Materia. Departamento F\'{\i}sica 
  Molecular, CSIC, Serrano 121, E--28006 Madrid, Spain.} 
\email{cerni@astro.iem.csic.es}
\and
\author{Yves Benilan}
\affil{Laboratoire Interuniversitaire des Syst\`{e}mes Atmosph\'{e}riques,
 UMR 7583 CNRS, Universit\'{e}s Paris VII-XII, 94010 Cr\'{e}teil CEDEX,
 France.}

\altaffiltext{1}{Visiting Scientist, Division of Physics, Mathematics and Astronomy, 
California Institute of Technology, MS 320-47, Pasadena, CA 91125, USA}

\begin{abstract}
We report on the detection with the Infrared Space Observatory (ISO)
of a molecular band at 57.5 $\mu$m (174 cm$^{-1}$)
in carbon-rich evolved stars and in Sgr B2. Taking into account the
chemistry of these objects the most likelihood carrier is a
carbon chain. We tentatively assign the band to the $\nu_5$ bending mode of C$_4$
for which a wavenumber of 170-172.4 cm$^{-1}$ has been derived in matrix
experiments (Withey et al. 1991).
An alternate carrier might  be C$_4$H, although  the frequency
of its lowest energy vibrational bending mode, $\nu_7$, is poorly known
(130-226 cm$^{-1}$).
If the carrier is C$_4$, the derived maximum abundance 
is nearly similar to that found for C$_3$ in the interstellar
and circumstellar media by Cernicharo et al. (2000). Hence, 
tetra-atomic carbon could be one of the most abundant carbon chain molecules
in these media.

\end{abstract}


\section{Introduction}
Long carbon chain radicals have been detected in interstellar and
circumstellar clouds through their pure
rotational spectrum at radio wavelengths (C$_5$H, C$_6$H, C$_7$H,
C$_8$H, H$_2$C$_3$, H$_2$C$_4$: Cernicharo et al. 1986, 1987,
Gu\'elin et al. 1987, 1997, Cernicharo and Gu\'elin 1996).
Polyynes such as  C$_2$H$_2$, C$_4$H$_2$, C$_6$H$_2$ and single aromatic
species such as  benzene, have been detected in the protoplanetary nebula
CRL618 (Cernicharo et al. 2001a\&b) supporting the idea that much
complex carbon-rich molecules could be formed in space.
However, the mechanisms allowing the growth
of carbon-rich molecules are still poorly known and the full set
of molecules that could participate in the chemical
reactions leading to the formation of large complex carbon-rich
species has yet to be identified. Among these "building blocks" the
C$_n$ chains have attracted the interest of laboratory
spectroscopists and astronomers.
C$_3$ has been observed at optical wavelengths in
the atmospheres of cool stars (see, e.g., Zuckerman et al. 1976),
and it has been identified in the envelope of IRC+10216 through
its $\rm \nu_3$ antisymmetric stretching mode in the mid-infrared
by Hinkle et al. (1988) who derived $\chi$(C$_3$)$\simeq$10$^{-6}$.
This key species has recently been found in the ISM (Cernicharo et al.
1996, 1997a, 2000, Giesen et al. 2001) through its $\nu_2$
low frequency bending mode (Schmuttenamer et al. 1990)
and in the diffuse interstellar clouds through
its electronic transition near 4052 \.{A} (Maier et al. 2001; Roueff
et al. 2002).
It has been suggested that C$_3$
could be involved in the formation of the diffuse interstellar bands
(Douglas, 1977; Clegg and Lambert 1982).
ISM observations in Cernicharo et al. (2000) put a
lower limit of 10$^{-8}$ to the C$_3$ abundance.
In addition, C$_5$ has been detected in  the C-rich star IRC+10216 with an
abundance ratio C$_3$/C$_5\simeq$10 (Bernath et al. 1989).
Less is known about numbered carbon chains, but C$_4$ could well be a very
abundant molecule both in C-rich evolved stars and in the ISM.

In this Letter we report the detection of a blend of lines at 57.5
$\mu$m that we tentatively assign to the $\nu_5$ bending mode of C$_4$.
The feature has been observed in Sgr B2, IRC+10216, CRL618, CRL2688,
and NGC7027. We discuss the possibility that the carrier could be
another different carbon chain : C$_4$H, although
photodetachement experiments
suggest that its lowest energy bending mode could be at higher
frequencies (Taylor et al. 1998).

\section{Observations and Results}
We have searched for the emission/absorption of the low energy bending
modes of polyatomic molecules
toward interstellar and circumstellar clouds using the Long Wavelength
spectrometer (LWS; Clegg et al. 1996, Swinyard et al. 1996)
on board ISO$^1$ (Kessler et al. 1996). Our main target for this
search has been the C-rich evolved star IRC+10216.

The LWS grating data of IRC+10216 were analyzed 
and modeled by Cernicharo et al. (1996). The data have been reprocessed
following the last pipeline product (TDT 19800158). No significant
differences have been found with the already published analysis. 
After identifying all the HCN (including several vibrational states) and
CO lines, a conspicuous feature composed by a blend of several lines (see 
Figure 1, top panel) was identified between 56 and 58 $\mu$m. 
These  lines could not be assigned to any known species and
no other similar patterns were found in the ISO far-infrared
spectrum of IRC+10216.
We have checked for all pure rotational lines
of light species that could be potentially abundant 
without success. Hence, the only possible explanation is that
the feature corresponds to a bending mode of an abundant heavy
molecule, and most likely to a carbon chain.
We searched for the same feature toward other objects
selected from the same criteria of 
our previous searches of C$_3$, C$_4$H$_2$, C$_6$H$_2$, and benzene
(Cernicharo et al. 1996, 1997a, 2000, 2001a,b), i.e., IRC+10216, CRL618, 
CRL2688, NGC7027, and Sgr B2 (among other sources). 

For Sgr B2 M we have used the LWS grating spectra obtained during revolutions
287 and 494 (TDTs 28701401, 287012130, 287012131, and 49400302). Some of
these data have been already presented (Cernicharo et al. 1997b, 
Goicoechea \& Cernicharo 2001).
All together represent a total of 32 individual scans. The feature at
57.5 $\mu$m is clearly present. There are many other LWS grating
observations of Sgr B2 with a reduced number of scans. We have inspected all
of them and, again, the feature appears in each individual
scan. Sgr B2 N  was observed in the course of a fast mapping of the Sgr B2
star forming region with the LWS spectrometer and only four scans are available
(see Cernicharo et al. 1997b).

For the Planetary Nebula NGC7027 we have used all the LWS grating data taken by
ISO (see Herpin et al. 2002 for details). The data for the Proto-Planetary
Nebula (PPN) CRL618 (TDTs 68800302 and 68800450) are those published by
Herpin and Cernicharo (2000).
For CRL2688, a young PPN, the TDT 02101504 data have been used. A previous
analysis of the ISO/LWS spectrum of this source has been presented by Cox et
al. (1996).

All the spectra have been analyzed using the ISO spectrometers
data reduction package ISAP$^{2}$.
The data products correspond 
to version 10 of the
pipeline. A polynomial baseline had to be removed from the final spectrum
in each source.
For some objects this baseline removal is  
critical to correctly extract 
the shape and intensity of the absorption features.
Due to the large number of lines in Sgr B2 M and N spectra,
the determination of the continuum level is not obvious.
For sources with emission lines, although baseline removal is less
critical, the global behavior of detector SW2 does not follow the expected
shape for continuum emission and a polynomial of fourth degree has
been removed after blanking the wavelength intervals where features above
5 sigma were present.

\footnotetext[1]
{ Based on observations with ISO, 
an ESA project with instruments funded by ESA Member States 
(especially the PI countries: France, Germany, the Netherlands 
and the United Kingdom) and with participation of ISAS and NASA.} 
\footnotetext[2]
{The ISO Spectral Analysis Package (ISAP) is a joint development by the LWS 
and SWS Instruments Teams and Data Centers. Contributing institutes are CESR, IAS,
 IPAC,MPE, RAL, and SRON.}

Figure 1 shows the 53-62 $\mu$m spectra of the selected sources. We have
also analyzed the available observations for several other sources : O-rich
stars, Orion, Sgr A, bright molecular sources, and even Mars to check for possible
instrumental effects.
The new infrared feature was not detected in any of these sources.
We have also examined the spectral response
of the LWS-SW2 filter in order to check for possible instrumental
contributions to the blend of lines found around 57 $\mu$m but
there is nothing at this wavelength (Garcia--Lario, private
communication). Hence, we conclude that  the blend of lines is real.
The pure
rotational lines of CO and HCN are indicated in the top panel of
Figure 1. The model of Cernicharo et al. (1996) predicted a low
flux for these high-J lines. Only in IRC+10216 some of
the CO lines (J=43-42, 44-43 and 45-44) are marginally detected
at a 3$\sigma$ level.

The observed band consists of a blend of at least 4 spectral lines or
sub-bands (at 57.0, 57.6, 57.9 and 58.2 $\mu$m).
There is an additional feature  separated
from the band at 56.15 $\mu$m that is detected in C-rich
evolved stars but is less evident in Sgr B2 N and M.
The J=47-46 line of CO could only make a modest contribution to this
feature in C-rich evolved stars since other high-J CO lines are weak --
even in IRC+10216. In Sgr B2 M 
this feature is blended with the absorption produced by
NH$_3$ (J=9-8). The 56.15 $\mu$m feature could be one of the fine
structure lines of [SI] at 56.31 $\mu$m ($^3$P$_0$--$^3$P$_1$) but 
its companion [SI] ($^3$P$_1$--$^3$P$_2$) line  at 25.249 $\mu$m is missing
in all objects.
The  feature could also be the R(1) line of HD. However, it is
unlikely to find such a high intensity for this line in evolved stars.
We think, that if instrumental, the feature  could arise from Uranus, the
planet used as calibrator for the LWS/grating spectrometer. However, its
intensity changes by more than a factor 10 between NGC7027 and
IRC+10216 or CRL618 (the continuum flux
at 58 $\mu$m is nearly the same for NGC7027 and
CRL618, 1.15 and 0.7~10$^{-16}$ W cm$^{-2}\mu$m$^{-1}$ respectively, and
4 and 3 10$^{-16}$ W cm$^{-2}\mu$m$^{-1}$ for IRC+10216 and CRL2688
respectively).
Therefore, 
the 56.2 $\mu$m feature in IRC+10216, CRL618, NGC7027 and CRL2688
cannot be explained alone by a possible instrumental contribution from
the calibrator. Figure 1 shows that the 56.2 $\mu$m line follows
the intensity
of the 57-58 $\mu$m features. In the following we will consider
that all them belongs to the same molecular band.

Two additional features at 53.7 and 54.4 $\mu$m in the spectrum of
IRC+10216 (labelled U1 and U2 in the top panel of Figure 1), and
another one at 52.89 $\mu$m (not shown), are
also detected in NGC7027 but not in the other sources.
These three lines remain, so far, unidentified.
The contribution of OH (Goicoechea \& Cernicharo 2002) and the maximum
possible contribution from H$_2$O (Cernicharo et al. 2002, in preparation)
to the spectra of Sgr B2 N and M are indicated in the
panel corresponding to Sgr B2 M. The contribution of H$_2$O to
the spectrum of CRL618 at 58$\mu$m is negligible (Herpin
and Cernicharo 2000).

\section{Discussion}
The fact that the 58 $\mu$m feature is detected in all the selected C-rich
post-AGB objects clearly points toward a carbon-rich molecule as the
carrier of the band.
Several molecules with four atoms have low frequency
bending modes around 160-220 cm$^{-1}$.
HC$_3$N has a low energy bending mode, $\nu_7$, at 222.413 cm$^{-1}$
($\simeq$45 $\mu$m).
Dyacetylene, C$_4$H$_2$, recently detected in CRL618 with a very large
abundance (see Cernicharo et al. 2001a\&b), has a low energy bending mode,
$\nu_9$, also at $\simeq$45 $\mu$m. Another example of molecule with
bending modes in this wavelength range is C$_4$H.
Pure rotational lines from the $\nu_7=1,2$
bending states of C$_4$H have been detected in IRC+10216 by Gu\'elin et
al. (1987) and Yamamoto et al. (1987). The frequency of this mode
is expected from ab initio calculations (Graf et al. 2001) to be around 
178 cm$^{-1}$.
Yamamoto et al. (1987) have estimated a frequency of 131 cm$^{-1}$ from
the l-doubling constant, q$_7$, of the $\nu_7$=1,2 vibrational levels.
A 20\% of error could be expected on this value  (Mikami et al. 1989).
Photoelectron spectra of C$_4$H indicated, however, a
larger frequency for this mode (226 cm$^{-1}$; Taylor et al. 1998).
The low lying $^2\Pi$ electronic state of C$_4$H is at  $\sim$468 cm$^{-1}$
above the ground (Taylor et al. 1998).
Both transitions, the $\nu_7=1-0$ and
the electronic $A^2\Pi-X^2\Sigma$ are in the wavelength coverage
of the Short Wavelength Spectrometer (SWS) on-board ISO.
However, IRC+10216 is so bright in the mid-infrared
that the expected range for these transitions is covered by strong ripples
and not obvious features have been found so far (Cernicharo et al. 1999).
The shape of the band in Figure 1 allows to rule out a
$^1\Pi-^1\Sigma$ transition, i.e., the bending modes of HC$_3$N, HC$_5$N,
C$_n$H$_2$ with n even,
C$_n$ with n odd, etc. These molecules have a $^1\Sigma$ ground electronic
state and, at the limited resolution of the LWS grating spectrometer, a
strong Q-branch with weak P and R-branches are expected for their bending
modes. The HC$_3$N $\nu_7$ and C$_4$H$_2$ $\nu_9$ modes are present in the
LWS spectrum of CRL618 (to be published elsewhere), and correspond to
these expectations.

Taking into account the large abundance of C$_3$ found
in Sgr B2 and IRC+10216, C$_4$ could be a good candidate to explain
the  IR feature under investigation.
Linear C$_4$ was first observed in the laboratory
using electron spin resonance spectroscopy of carbon clusters in
matrices by Graham et al. (1976). They
were able to derive the spin-coupling and spin-doubling constants,
$\lambda$ and $\gamma$, for the ground vibrational state.
C$_4$ was first detected and characterized in the gas phase
by Heath and Saykally (1991) by  observing several ro-vibrational lines
of the $\nu_3$ antisymmetric stretching mode.
Moazzen-Ahmadi et al. (1994)
observed the same band and the $\nu_3+\nu_5-\nu_5$ hot band and determined
the rotational constants, B and D, for both the ground and the $\nu_5$ states.
The l-doubling constant, q$_5$, for the latter vibrational state was also
derived leading these authors  to estimate a frequency of 160$\pm$4 cm$^{-1}$
for the $\nu_5$ mode.
The $\nu_5$ bending mode of C$_4$ has been measured in matrix experiments
at a wavelength of 58 $\mu$m (Withey et al. 1991), very
close to that of the newly found IR feature.
They observed two different lines separated by 2 cm$^{-1}$ (170 and 172.4
cm$^{-1}$) that they interpreted as two different trapping
sites for C$_4$ in the Ar matrix. Both  values could correspond
to the frequency of the $\nu_5$ mode of C$_4$ but the authors adopted
the value of 172.4 cm$^{-1}$. This value agrees well with the series
of peaks observed in the photoelectron spectroscopy of Arnold et al. (1991).
The difference between the frequency of the new IR band and that derived
from the matrix experiments is 2.6 cm$^{-1}$, which could be consistent
with a frequency shift introduced by the Ar matrix of 1.5\%.
Ab initio calculations predict wavenumbers
in the 175-214 cm$^{-1}$  range (47-57 $\mu$m)
and an infrared intensity
of 46.2 km mol$^{-1}$ [A($\nu_5=1-0$) $\simeq$ 0.25 s$^{-1}$;
Martin et al. 1991;
Watts et al. 1992; Taylor and Martin, 1996; see also Van Orden
and Saykally, 1998]. 
Due to the  complexity of the C$_4$ electronic energy
levels and to the limited accuracy level of ab initio
calculations, this intensity value has to be taken into account
as a crude estimation.

The fact that the observed
band (see Figure 1) contains several spectral features suggests a
transition between terms of higher spin multiplicity, in particular toward
a vibronic transition $^2\Pi-^2\Sigma$ or  $^3\Pi-^3\Sigma$, i.e.,
just the type of bands we can expect for C$_4$H and  
C$_4$ which have  $^2\Sigma$ and $^3\Sigma_g^-$ ground electronic states
respectively. Consequently the lowest bending modes of these species,
$\nu_7$ for C$_4$H and $\nu_5$ for C$_4$, will have
$^2\Pi$ and $^3\Pi$ vibronic character respectively.
An argument against
C$_4$H is the following,
while pure rotational lines of its bending mode
have been detected towards IRC+10216 (Gu\'elin et al. 1987; 
Yamamoto et al. 1987) these lines are missing in all other objects
reported here (except in CRL618 where several weak lines from these
states have been detected; Cernicharo et al. 2002, in preparation).
Moreover, the abundance of
C$_4$H in Sgr B2 is rather low (Gu\'elin, Mezaoui \& Friberg, 1982),
while C$_3$ has a large abundance
(Cernicharo, Goicoechea and Caux, 2000).
Hence, it is tempting to assume that the observed features in post-AGB 
stars and in 
the ISM correspond to the $\nu_5$ mode of C$_4$, although a definitive
assignment will require better spectral resolution data and/or
additional laboratory information on this species.
The main problem in assigning the  band to the $\nu_5$ mode of
C$_4$ is the
large spin-orbit interaction constant, A$_{SO}$, that should have  
the $^3\Pi$ vibronic state in order to reproduce the observations. As the 
ground electronic state of C$_4$ is $^3\Sigma$, a very small
value for A$_{SO}$ should be expected. However, if the molecule has a low
lying excited electronic state with $\Lambda\#0$, then the ground and
the excited states will be strongly coupled through vibronic mixing
and a large spin-orbit constant could be expected. 
This is the case for CCH which has a $^2\Sigma$
ground state and an A$^2\Pi$ state 3600 cm$^{-1}$ above. The $\nu_2$ bending
mode of the ground electronic state has A$_{SO}\simeq$0.3 cm$^{-1}$. For 
C$_4$H, A$_{SO}$ is very
large, $\simeq$3 cm$^{-1}$ (Yamamoto et al. 1987), because its A$^2\Pi$ state 
is only $\simeq$468 cm$^{-1}$ above the $^2\Sigma$ ground state (see above).
For C$_4$ the situation  could be very similar to that of C$_4$H as its
first excited electronic state is $ ^1\Delta$ and is only 2680 cm$^{-1}$
above the ground electronic $^3\Sigma$ state
(Arnold et al. 1991; Xu et al. 1997).
Last panel in figure 1
shows the computed band shape for A$_{SO}$=4 cm$^{-1}$ at
different temperatures.
The similarity between the observed and computed shape is appealing.
Nevertheless, we note that the adopted spin-orbit constant is
particulary large. We have run several models with different A$_{SO}$
values and conclude that the shape of the band is only crudely reproduced for
A$_{SO}>$2 cm$^{-1}$.
Another possibility that may explain the complexity of the band shape
and allows smaller A$_{SO}$ values is that  $^{13}$CCCC and C$^{13}$CCC 
isotopomers also make a remarkable contribution
to the observed emission. 
This could also apply to C$_4$H if finally
this molecule is confirmed as the carrier. We note, however,
that in this case we should also expect a very large column
density of C$_4$H, which looks incompatible with the upper limits found
for this molecule in Sgr B2 (Gu\'elin, Friberg and Mezaoui 1982).
If the carrier is C$_4$ then the column density will be
particularly large in all observed sources. Assuming
a vibrational excitation temperature of 50-100 K in AGB stars and PPNs, 
N(C$_4$) has to be close to 10$ ^{15}$ cm$^{-2}$, i.e., a typical abundance 
of a few 10$^{-7}$ - 10$ ^{-6}$.
In Sgr B2 the computed column density (assuming a lower
vibrational excitation temperature) is 1.2 10$^{15}$ cm$^{-2}$, i.e., similar
to that derived by Cernicharo, Goicoechea and Caux (2000) for C$_3$.
The accuracy of the band intensity from ab initio calculations dominates the
error in the estimated column densities. 

Future space--born platforms, such as the Herschel Space Observatory, will be
equipped with heterodyne instruments with much better spectral
resolution which should permit the detection of longer C$_n$ chains 
through their low lying bending modes in the far--IR. 
The detection of these molecules and the determination of their
abundances will allow to stablish their role in the formation 
of the carriers of the unidentified infrared bands.
Waiting for Herschel, important
work has to be done in  spectroscopy laboratories to fully characterize
these molecular species.

\acknowledgments
We thank Spanish DGES and PNIE for funding
support under grant ESP98--1351--E and PANAYA2000-1784.
JRG acknowledges \textit{Universidad Aut\'onoma de Madrid} for a pre--doctoral
fellowship. We would like to thank M. Gu\'elin and J.R. Pardo for their
useful comments and suggestions.

\clearpage

\begin{figure}
\caption{Observed ISO/LWS spectra toward the sources discussed in the text.
The wavelengths of the CO lines J=43-42
up to J=49-48 and those of HCN J=56-55 up to 64-63 are indicated by
arrows in the top panel. For IRC+10216, CRL2688, CRL618 and NGC7027 the continuum
level has been removed. For Sgr B2 N and Sgr B2 M the line
over continuum flux ratio is shown. In all cases the continuum has
been derived from a polynomial fit to the spectra after removing
all emission/absorption features above 3 $\sigma$. The bottom
panel shows the expected band shape for a transition $^3\Pi-^3\Sigma$
with A$_{SO}$=4 cm$^{-1}$, intrinsic linewidth of 10 kms$^{-1}$ and spectral resolution,
$\lambda/\Delta\lambda$, of 200.
}
\end{figure}

\plotone{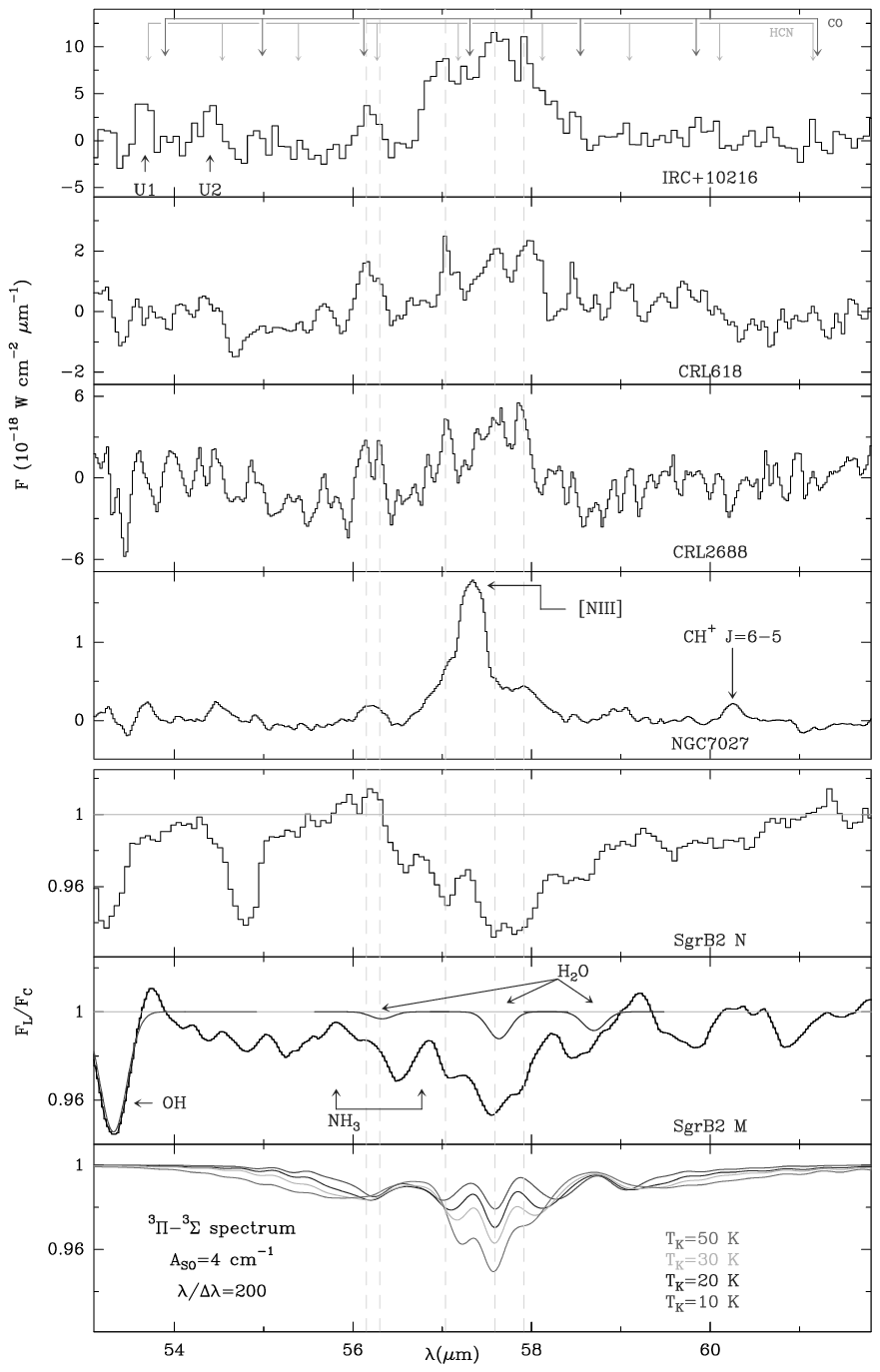}

\end{document}